\documentstyle[onecolumn,epsfig]{mn}

\newif\ifAMStwofonts
\def \eps{\epsilon}
\def \veps{\varepsilon}
\def \d{\partial}
\def \lam{\lambda}

\renewcommand{\(}{\left(}
\renewcommand{\)}{\right)}
\newcommand{\df}[2]{ \frac{\partial {#1}}{\partial {#2}} }

\newcommand{\I}{\mbox{\rm i}}

\ifoldfss
  \ifCUPmtlplainloaded \else
    \NewTextAlphabet{textbfit} {cmbxti10} {}
    \NewTextAlphabet{textbfss} {cmssbx10} {}
    \NewMathAlphabet{mathbfit} {cmbxti10} {} % for math mode
    \NewMathAlphabet{mathbfss} {cmssbx10} {} %  "   "    "
  \fi
  \ifAMStwofonts
    \ifCUPmtlplainloaded \else
      \NewSymbolFont{upmath} {eurm10}
      \NewSymbolFont{AMSa} {msam10}
      \NewMathSymbol{\upi}     {0}{upmath}{19}
      \NewMathSymbol{\umu}     {0}{upmath}{16}
      \NewMathSymbol{\upartial}{0}{upmath}{40}
      \NewMathSymbol{\leqslant}{3}{AMSa}{36}
      \NewMathSymbol{\geqslant}{3}{AMSa}{3E}

       \let\ge=\geqslant
    \fi
  \fi
\fi % End of OFSS

\ifnfssone
  \newmathalphabet{\mathit}
  \addtoversion{normal}{\mathit}{cmr}{m}{it}
  \addtoversion{bold}{\mathit}{cmr}{bx}{it}
  \newmathalphabet{\mathbfit} % math mode version of \textbfit{..}
  \addtoversion{normal}{\mathbfit}{cmr}{bx}{it}
  \addtoversion{bold}{\mathbfit}{cmr}{bx}{it}
  \newmathalphabet{\mathbfss} % math mode version of \textbfss{..}
  \addtoversion{normal}{\mathbfss}{cmss}{bx}{n}
  \addtoversion{bold}{\mathbfss}{cmss}{bx}{n}
  \ifAMStwofonts
    \ifCUPmtlplainloaded \else
      \UseAMStwoboldmath
      \makeatletter
      \new@mathgroup\upmath@group
      \define@mathgroup\mv@normal\upmath@group{eur}{m}{n}
      \define@mathgroup\mv@bold\upmath@group{eur}{b}{n}
      \edef\UPM{\hexnumber\upmath@group}
      \new@mathgroup\amsa@group
      \define@mathgroup\mv@normal\amsa@group{msa}{m}{n}
      \define@mathgroup\mv@bold\amsa@group{msa}{m}{n}
      \edef\AMSa{\hexnumber\amsa@group}
      \makeatother
      \mathchardef\upi="0\UPM19
      \mathchardef\umu="0\UPM16
      \mathchardef\upartial="0\UPM40
      \mathchardef\leqslant="3\AMSa36
      \mathchardef\geqslant="3\AMSa3E

       \let\ge=\geqslant
    \fi
  \fi
\fi % End of NFSS release 1

\ifnfsstwo
  \DeclareMathAlphabet{\mathbfit}{OT1}{cmr}{bx}{it}
  \SetMathAlphabet\mathbfit{bold}{OT1}{cmr}{bx}{it}
  \DeclareMathAlphabet{\mathbfss}{OT1}{cmss}{bx}{n}
  \SetMathAlphabet\mathbfss{bold}{OT1}{cmss}{bx}{n}
  \ifAMStwofonts
    \ifCUPmtlplainloaded \else
      \DeclareSymbolFont{UPM}{U}{eur}{m}{n}
      \SetSymbolFont{UPM}{bold}{U}{eur}{b}{n}
      \DeclareSymbolFont{AMSa}{U}{msa}{m}{n}
      \DeclareMathSymbol{\upi}{0}{UPM}{"19}
      \DeclareMathSymbol{\umu}{0}{UPM}{"16}
      \DeclareMathSymbol{\upartial}{0}{UPM}{"40}
      \DeclareMathSymbol{\leqslant}{3}{AMSa}{"36}
      \DeclareMathSymbol{\geqslant}{3}{AMSa}{"3E}

       \let\ge=\geqslant
    \fi
  \fi
\fi % End of NFSS release 2

\ifCUPmtlplainloaded \else
  \ifAMStwofonts \else % If no AMS fonts
    \def\upi{\pi}
    \def\umu{\mu}
    \def\upartial{\partial}
  \fi
\fi

\title[On the $r$-mode spectrum of relativistic stars: Inclusion of the radiation reaction]{On the $r$-mode spectrum of relativistic stars: Inclusion of the radiation reaction}

\author[J.~Ruoff and K.D.~Kokkotas]
{Johannes Ruoff and Kostas D.~Kokkotas \\
  Department of Physics, Aristotle University of Thessaloniki,
  Thessaloniki 54006, Greece}

\date{Accepted ???? Month ??.
      Received 2001 Month ??;
      in original form 2001 Month ??}

\pagerange{\pageref{firstpage}--\pageref{lastpage}}
\pubyear{2001}
\begin{document}

\maketitle

\label{firstpage}

\begin{abstract}
  We consider both mode calculations and time evolutions of axial
  $r$-modes for relativistic uniformly rotating non-barotropic neutron
  stars, using the slow-rotation formalism, in which rotational
  corrections are considered up to linear order in the angular
  velocity $\Omega$. We study various stellar models, such as uniform
  density models, polytropic models with different polytropic indices
  $n$, and some models based on realistic equations of state.  For
  weakly relativistic uniform density models, and polytropes with
  small values of $n$, we can recover the growth times predicted from
  Newtonian theory when standard multipole formulae for the
  gravitational radiation are used. However, for more compact models,
  we find that relativistic linear perturbation theory predicts a
  weakening of the instability compared to the Newtonian results. When
  turning to polytropic equations of state, we find that for certain
  ranges of the polytropic index $n$, the $r$-mode disappears, and
  instead of a growth, the time evolutions show a rapid decay of the
  amplitude. This is clearly at variance with the Newtonian
  predictions. It is, however, fully consistent with our previous
  results obtained in the low-frequency approximation.
\end{abstract}


\section{Introduction}

The $r$-mode instability, and the associated spin-down of rapidly
rotating young neutron stars, has created considerable interest in the
astrophysical community, particularly as this may be a detectable
source of gravitational waves. However, it was soon realized that a
lot of effects tend to work against the growth of the $r$-mode.
Viscous damping, coupling to the crust, magnetic fields and
differential rotation could severely reduce the importance of the
instability, if not suppress it completely.  Since there is a large
number of papers published on this subject we refer the reader to
two recent reviews (Andersson \& Kokkotas, 2001; Friedman \& Lockitch,
2001).

Most of the results have been obtained in a Newtonian context, with
radiation effects incorporated through the standard multipole
formulae.  However, in a first step towards a fully relativistic
treatment, it was first shown by Kojima (1997, 1998) that the
Newtonian treatment might actually miss important relativistic
effects, such as the frame dragging. Working in the low-frequency
approximation, which actually neglects any gravitational radiation, he
showed that instead of having a discrete $r$-mode frequency, the frame
dragging, being a function of the stellar radius $r$, gives rise to a
whole band of frequencies. The existence of this continuous spectrum
was proven by Beyer \& Kokkotas (1999) in a rigorous way. However, it
is still possible to find mode solutions for uniform density stars, as
has been shown by Lockitch, Andersson \& Friedman (2001). In a
previous paper (Ruoff \& Kokkotas, 2001), which we refer to henceforth
as paper I, we have corroborated these results both through explicit
mode calculations and through the evolution of the time dependent
equations. However, at the same time we have shown that certain
polytropic models, and some stellar models based on realistic
equations of state, do not allow the existence of physical $r$-modes.
Although for $l=2$, it was always possible to find a mathematically
valid mode solution with its eigenfrequency lying inside the
continuous band, the associated fluid velocity was diverging at some
point inside the star. Therefore, we considered such solutions as
being unphysical, and hence non-existent. This view was supported by
the explicit time evolution of the relevant equations, as the system
did not show any tendency to oscillate with the ``unphysical''
frequency predicted by the mode calculation.

These results were obtained in the low-frequency approximation, which
amounts to neglecting the radiative metric perturbations. In this
paper, we consider the complete set of axial equations, which follow
from expanding the Einstein equations up to first order in the
rotation rate $\Omega$. However, we do not include the coupling to the
polar equations. As we shall show, the main results from paper I
concerning the existence of the $r$-mode more or less directly carry
over to the fully relativistic case. This means that even if one takes
into account the radiative back-reaction, there are certain cases
where, at least within the slow-rotation approximation, discrete
purely axial $r$-modes cannot be found.

\section{Mathematical formulation}

If we focus on purely axial perturbations, the perturbed metric can be
written in the following form:
\begin{equation}\label{metric}
  ds^2 = ds^2_0 + 2\sum_{l,m}\(h_0^{lm}(t,r)dt + h_1^{lm}(t,r)dr\)
  \(-\frac{\d_\phi Y_{lm}}{\sin\theta}d\theta
  + \sin\theta\d_\theta Y_{lm}d\phi\)\;,
\end{equation}
where $Y_{lm} = Y_{lm}(\theta,\phi)$ denote the scalar spherical
harmonics. The unperturbed metric $ds^2_0$ represents a non-rotating
stellar model with a first order rotational correction $\omega$ (the
frame dragging) in the ${t\phi}$-component:
\begin{equation}
  ds^2_0 = -e^{2\nu} dt^2 + e^{2\lam} dr^2
  + r^2\(d\theta^2 + \sin^2\theta d\phi^2\)
  - 2\omega r^2\sin^2\theta dtd\phi\;.
\end{equation}
The fluid is assumed to rotate with uniform angular velocity
$\Omega$.  The axial component of the fluid velocity perturbation can
be expanded as
\begin{equation}\label{fluid}
  4\pi(p + \eps)\(\delta u^\theta, \delta u^\phi\)
  = e^{\nu}\sum_{l,m}U^{lm}(t,r)
  \(-\frac{\d_\phi Y_{lm}}{\sin\theta}, \sin\theta\d_\theta Y_{lm}\)\;.
\end{equation}
Einstein's field equations then reduce to three equations for the
three variables $h_0^{lm}$, $h_1^{lm}$ and $U^{lm}$ (Kojima 1992).

To obtain a system of equations, which is suitable for the numerical
time evolution, the most straightforward way is to use the
ADM-formalism (Arnowitt, Deser \& Misner 1962, Ruoff 2001).  For the
axial perturbations, the expansion of the metric variables $V_4$, and
extrinsic curvature variables $K_3$ and $K_6$, has been given in paper
I. We just restate the connection between those variables and the ones
defined above:
\begin{eqnarray}
  h_0 &=& e^{\nu-\lam} K_6\;,\\
  h_1 &=& e^{\lam-\nu} V_4\;,\\
  U &=& 4\pi e^{-\lam-\nu}(p + \eps)\(u_3 - K_6\)\;.
\end{eqnarray}
The evolution equations for the variables $V_4$, $K_3$, $K_6$ and
$u_3$ are
\begin{eqnarray}
  \label{V4}
  \(\df{}{t} + \I m\omega\)V_4 &=& e^{2\nu-2\lam}
  \left[K_6' + \(\nu' - \lam' - \frac{2}{r}\)K_6 - e^{2\lam}K_3\right]\;,\\
  \label{K3}
  \(\df{}{t} + \I m\omega\)K_3 &=& \frac{l(l+1) -2}{r^2}V_4
  + \frac{2\I m}{l(l+1)}\omega'e^{-2\lam}K_6\;,\\
  \label{K6}
  \(\df{}{t} + \I m\omega\)K_6 &=& V_4' - \frac{\I mr^2}{l(l+1)}
  \left[\omega'K_3 - 16\pi\varpi(p + \eps)u_3\right]\;,\\
  \label{u3}
  \(\df{}{t} + \I m\Omega\)u_3 &=& \frac{2\I m\varpi}{l(l+1)}
  \(u_3 - K_6\)\;,
\end{eqnarray}
with the abbreviation
\begin{equation}
  \varpi = \Omega - \omega\;.
\end{equation}
Furthermore, we have one momentum constraint:
\begin{equation}\label{MC_odd}
  16\pi(p + \eps)u_3 = K_3' + \frac{2}{r}K_3
  - \frac{l(l+1) - 2}{r^2}K_6
  - \frac{2\I m\omega'}{l(l+1)}e^{-2\nu}V_4\;.
\end{equation}
We could use this constraint to replace $u_3$ in Eq.~(\ref{K6}) and
thus obtain a complete set of three equations. However, we prefer to
use the above set of four equations and use the constraint to monitor
the numerical evolution. Of course, we also need the constraint
(\ref{K6}) to construct valid initial data. We evolve the above system
using a two-step Lax-Wendroff scheme yielding second order
convergence, which can be easily checked by monitoring the violation
of the momentum constraint (\ref{MC_odd}) for different spatial
resolutions.

The above system of evolution equations describe both the $w$- and
$r$-modes. In this paper, we would like to focus on the $r$-modes
only, and so we will try to minimise the $w$-mode contribution by
choosing appropriate initial data. In addition we will compare the
evolution of the full system with the evolution of the equations in
the low-frequency approximation.

In the low-frequency approximation, $V_4 = 0$, and the evolution
system involves only the variables $K_6$ and $u_3$. As initial data,
we chose an arbitrary function for $u_3$ and solved the elliptic
equation (32) of paper I for $h_0 = e^{\nu-\mu}K_6$. For the full
system, we can do exactly the same. We set $V_4(t=0) = 0$ and choose
an arbitrary function for $u_3$. Next, we solve Eq.~(32) of paper I
for $K_6$ and finally compute the remaining variable $K_3$ from the
momentum constraint (\ref{MC_odd}).

The conversion of the evolution system (\ref{V4}) -- (\ref{u3}) into a
time independent form is straightforward. Assuming a harmonic time
dependence $e^{-i\sigma t}$, we can deduce the following two ordinary
differential equations:
\begin{eqnarray}
  \label{V4ode}
  V_4' &=& \frac{\I mr^2}{l(l+1)}
  \(\omega'K_3 - 16\pi\varpi(p + \eps)u_3\)
  - \I\(\sigma - m\omega\)K_6\;,\\
  \label{K6ode}
  K_6' &=& e^{2\lam}K_3 - \(\nu' - \lam' - \frac{2}{r}\)K_6
  - \I e^{2\lam-2\nu}\(\sigma - m\omega\)V_4\;,
\end{eqnarray}
together with the two algebraic relations:
\begin{eqnarray}
  \label{K3alg}
  K_3 &=& \frac{\I}{\sigma - m\omega}\(\frac{l(l+1) -2}{r^2}V_4
  + \frac{2\I m\omega'}{l(l+1)}e^{-2\lam}K_6\)\;,\\
  \label{u3alg}
  u_3 &=& \frac{2m\varpi}
  {2m\varpi + l(l+1)(\sigma - m\Omega)}K_6\;.
\end{eqnarray}
Eq.~(\ref{u3alg}) is equivalent to Eq.~(40) of Paper I and is
singular if the denominator becomes zero, which can only happen if
$\sigma$ is real.  For real sigma, the expression $\sigma - m\omega$
can also vanish, and Eq.~(\ref{K3alg}) becomes ill defined. However,
as $\sigma$ is expected to be a complex frequency, both equations for
$K_3$ and $u_3$ should remain regular. Still, these singular
structures can pose difficulties for the numerical treatment when the
imaginary part of $\sigma$ becomes very small.

We should make a few more remarks on the above system. In principle we
could transform it into a single second order ODE for the quantity
$V_4$.  However, this equation would be rather messy since the
necessary algebra leads to many terms quadratic in $\Omega$ and
$\omega$. If, however, we neglect any terms quadratic or of higher
power in $\Omega$ and $\omega$, we can recover Eq.~(42) of Kojima
(1992) with the polar source terms $F_{l\pm1m}$ set to zero. This
equation was later used by Andersson (1998) to compute the rotational
corrections to the axial $w$-modes and the $r$-modes.

However, as we shall show, the values given by Andersson for the
$r$-modes cannot be corroborated neither through the time evolution
nor through mode calculation with the above set of equations. This is
because the extra terms that are not included in his form of the
equations are critical, at least for the study of the $r$-modes. To be
more specific, a basic property of the $r$-modes is that their
frequency $\sigma$ is proportional to the rotation rate $\Omega$. This
implies that the coefficient of $K_6$ in Eq.~(\ref{u3alg}) is not of
order $\Omega$ but of order 1, i.e.~$u_3$ and $K_6$ are of the same
order. If one looks for an equation similar to Eq.~(42) of Kojima
(1992), which is second order in $\sigma$, it is clear that one has to
retain all the $O(\Omega^2)$ terms if one wants to correctly compute
the $r$-mode frequencies, because in this case, it is $\sigma^2 =
O(\Omega^2)$. However, in Eqns.~(42) of Kojima (1992) and (10) of
Andersson (1998), the $O(\Omega^2)$ terms were discarded, hence these
equations are not suitable for computing correct $r$-mode frequencies.

Therefore, we leave Eqs.~(\ref{V4ode}) -- (\ref{K6ode}) as they are,
and solve them as a coupled first order system. To this end we follow
the method of Andersson (1998) and integrate (\ref{V4ode}) and
(\ref{K6ode}) using a complex valued $r$ coordinate in the exterior
spacetime.  In the asymptotic exterior, the behaviour of the
perturbation variables is given by
\begin{eqnarray}\label{asy}
  X \sim X_{\mbox{out}} e^{i\sigma r_*} + X_{\mbox{in}} e^{-i\sigma r_*}\;,
\end{eqnarray}
with $r_*$ being the well-known tortoise coordinate. The quasi-normal
modes are defined as purely outgoing solutions, i.e.~solutions with
$X_{\mbox{in}} = 0$. For a complex-valued frequency $\sigma$, the
amplitude of one of the terms in Eq.(\ref{asy}) will always diverge if
the integration is performed along the real $r$-axis. However, this
can be overcome by choosing the path of integration in the complex
$r$-plane to be a straight line with the angle to the real $r$-axis
given by arg($r$) = -arg($\sigma$). In a region far enough from the
stellar surface, we then test our solution by computing the Wronskian
with the pure outgoing solution $e^{i\sigma r_*}$. The vanishing of
the Wronskian tells us that we have found a quasi-normal mode of the
star. We have tested this method by computing the axial $w$-modes for
non-rotating neutron star models, reproducing the values given by
Kokkotas (1994) for uniform density models and other polytropic
models.

\section{Numerical results}

\subsection{Uniform density models}
In the Newtonian limit, one can estimate the growth time for the
$l = m = 2$ $r$-modes of uniform density models as (Kokkotas \&
Stergioulas, 1999)
\begin{equation}
  \label{g_rate}
  \tau_{PN} = 22\(\frac{1.4M_\odot}{M}\)
  \(\frac{\mbox{10 km}}{R}\)^4\(\frac{P}{\mbox{1 ms}}\)^6\mbox{s}\;.
\end{equation}
In general, the growth time is proportional to $P^{2l + 2}$, whereas
the oscillation period increases linearly with the rotation period
$P$. Of course, formula (\ref{g_rate}) has been derived by applying
the standard multipole formulae for the mass and current multipole
moments obtained from Newtonian theory. It is clear that for very
compact stars, there should be deviations due to higher order
corrections. In Fig.~1, we show the growth times for four uniform
density models with different values of the compactness $M/R =$ 0.1,
0.2, 0.3 and 0.4. In a double-logarithmic plot, a power law is
represented by a straight line. For the less relativistic stellar
models with $M/R =$ 0.1 and 0.2, the agreement of formula
(\ref{g_rate}) with the results of our mode calculations for $l=2$ is
very good, and for higher values of $l$, we can recover the $2l + 2$
dependence of the exponent, in fact we find that
\begin{equation}
  \tau_{PN} \sim (|m|\Omega)^{2l+2}\;.
\end{equation}
For more relativistic models, however, we find a systematic deviation
and the growth time rather goes with a power of $2l + 3$, for small
values of $P$. For large values of $P$, we still have a $2l + 2$
dependence, however, formula (\ref{g_rate}) tends to underestimate the
growth times, i.e.~the instability is not as strong as formula
(\ref{g_rate}) suggests. This weakening of the instability for very
compact stellar models might be linked to the formation of a potential
well inside the star, which can trap the gravitational waves. This
would make it more difficult for the mode to grow.  Nevertheless, for
uniform density models with a compactness in the range of realistic
neutron star models, the agreement between relativistic linear
perturbation theory and the results from the multipole formula is very
good.

In addition to mode calculations, we also performed the explicit
evolution of the time dependent equations. The results for the four
different compactness ratios are shown in Fig.~2. To make the growth
time accessible to the numerical evolution, we have chosen the
rotation rate to be $\veps = 2.0$ corresponding to values of $P
\approx 0.2\,$ms. This is significantly above the Kepler limit, which
is given approximately by $\veps = \Omega/\sqrt{M/R^3} \approx 0.7$.
Above this rotation rate the star becomes unstable with respect to
mass shedding.  Of course, in this case, the slow-rotation
approximation is not valid any more, but if implemented correctly, the
mode calculation and the time evolution should always yield the same
complex-valued frequencies for any given value of $\veps$. In all four
cases, it is clearly discernible that after a short initial time, an
exponentially growing mode appears, which soon dominates the
evolution. From the mode calculations we obtain the growth times given
in Table 1. In the graphs of Fig.~2, we have indicated these growth
times by straight lines, matching exactly the amplitudes of the
growing modes.
\begin{table}
  \caption{\label{models} The $r$-mode frequency $f$ and its growth
    time $\tau$ obtained from mode calculations for uniform density
    models with a central density of $\eps_c = 10^{15}$ g/cm$^3$ and
    rotational parameter $\veps = 0.5$. The last column contains the
    growth time $\tau_{PN}$ computed by formula (\ref{g_rate}).}
  \begin{tabular}{cccccccc}
    \rule[-2.5mm]{0mm}{7.5mm}
    $M/R$ & $M\;[M_\odot]$ & $R\;[$km$]$ & $p_c\;[$dyn/cm$^2]$ &
    Period [ms] & $f$[Hz] & $\tau$[s] &$\tau_{PN}$[s] \\
    \hline
    \rule[0mm]{0mm}{5mm}
    0.1 & 0.384 & 5.671 & $5.637\times10^{34}$ & 0.546 & 1830 & 150 & 139\\
    0.2 & 1.086 & 8.019 & $1.530\times10^{35}$ & 0.524 & 1906 & 13 & 12\\
    0.3 & 1.996 & 9.822 & $3.681\times10^{35}$ & 0.495 & 2022 & 3.3 & 3 \\
    0.4 & 3.072 & 11.342 & $1.454\times10^{36}$ & 0.449 & 2226 & 1.3 & 1.1\\
  \end{tabular}
\end{table}

\subsection{Polytropic models}

In paper I, we have shown that in the low-frequency approximation, not
all the polytropic stellar models can admit physical $r$-mode
solutions. Nevertheless, it was in most cases possible to find a
mathematically valid eigensolution satisfying the imposed boundary
condition. However, if the mode frequency $\sigma$ lay inside the
range of the continuous spectrum, the associated fluid velocity $u_3$
was diverging at some point inside the star. Moreover, the metric
variable $h_0$ exhibited an infinite slope at this particular point.
Based on these facts, we considered such solutions as being
unphysical, and hence nonexistent. We we able to corroborate our point
of view by explicitly evolving the time dependent equation, which
showed that there was no sign of a mode at the frequency predicted by
the mode calculation.  A further indication came from the behaviour of
the amplitude of the metric variable $h_0$.  In those cases where a
physical mode existed, the amplitude remained constant after a certain
initial time, whereas in the other cases, the amplitude kept decaying.

Of course, the low-frequency approximation is quite restrictive, since
it cuts off the gravitational radiation. This is the reason why the
associated eigenvalue problem becomes singular, as the mode
frequencies have to be real valued. With the inclusion of the
gravitational radiation, each mode acquires a finite imaginary part,
being positive for a damped mode and negative for a growing mode.  The
addition of an imaginary part (no matter how small) to the mode
frequencies removes the singular structure of the equations. Hence, it
has been argued that in considering the problem with the inclusion of
the gravitational radiation effect, one should be able to find the
$r$-modes, with the anticipated growth times derived from Newtonian
theory.

We find that this does not prove to be the case. For uniform density
models, the equations are {\it always} regular, i.e.~both with and
without the radiative terms, and it is always possible to find a
physical $r$-mode. However, all those stellar models which did not
admit physical $r$-modes in the low-frequency approximation {\it
  still} do not do so when radiation reaction is included.  In other
words, the qualitative picture which has emerged from the
low-frequency approximation does not change at all.

In Fig.~3, we show the time evolution of the same initial data for a
sequence of polytropic stellar models with fixed compactness $M/R =
0.3$, fixed rotation parameter $\veps = 1.0$, but varying polytropic
index $n$, ranging from $n=0$ to $n=1$. For $n=0$, 0.2 and 0.4, we
find the expected exponential growth, in agreement with the values
computed from explicit mode calculations. However, for the other
values $n=0.6$, 0.8 and 1.0, no growth is visible and instead the
amplitude decays. This decay becomes stronger as $n$ is increased.

In Fig.~4, we show the imaginary parts of the frequency for a sequence
of polytropic models with the polytropic index $n$ ranging from $n =
0.3$ to $n = 0.8$ in steps of 0.05. Both panels show the same data. In
the right panel, however, the plot is on a logarithmic scale and for a
larger range of $\veps$. It should be noted that the Kepler limit is
reached for $\veps \approx 0.7$, hence any value of $\veps$ above is
unphysical. We nevertheless show the behaviour of the growth times up
to $\veps = 2.5$, in order to gain more insight into how the mode
disappears. For $n = 0.3$, the growth times follows the expected power
law in the range from $\veps = 0.1$ to $\veps \approx 1.5$. For larger
$\veps$, a deviation becomes obvious. As $n$ is increased, this
deviation becomes more pronounced and a bulge starts to emerge at
$\veps \approx 1.8$, at which the imaginary parts tend to become
smaller. Recall that smaller imaginary part means longer growth time.
For $n = 0.5$, the curve exhibits a local minimum at $\veps \approx
1.8$. For $n \ge 0.55$ the curves cross the zero line at some specific
value of $\veps$ and reappear at some larger value of $\veps$.  The
larger $n$ is, the sooner the imaginary part becomes zero, and the
larger $\veps$ is at the reappearance point. For $n=0.8$, the
imaginary part becomes zero at $\veps\approx 0.6$, for larger $n$
this occurs even for much smaller $\veps$.

From the left panel of Fig.~4, one could be led to the conclusion that
the imaginary part does not vanish but actually crosses the
$\veps$-axis and becomes negative. This would correspond to the
$r$-mode being damped.  However, this seems not to be the case.  What
happens instead is that it becomes impossible to numerically find the
mode at all. Of course, numerically not finding the mode does not mean
that the mode ceases to exist for certain parameters, but from the
analogy with the results of the low-frequency approximation, we
conclude that the $r$-mode does indeed vanish, and there is only the
(non-radiative) continuous spectrum left. This view is corroborated by
the numerical evolutions, which in those cases show the decay of the
metric amplitude, whereas each fluid layer oscillates with its own
frequency, keeping a constant amplitude in exactly the same way as it
did in the low-frequency approximation. This is clarified in Fig.~5,
where we show the evolution of the same initial data, once using the
full set of equations, and once using the low-frequency approximation.
We plot the amplitudes of both the fluid variable $u_3$ (somewhere
inside the star) and the metric variable $K_6$ (somewhere outside the
star). As it can been seen, after about 10 ms, the fluid variable
$u_3$ remains at the same constant amplitude for both cases, whereas
$K_6$ keeps decaying. It decays even faster in the full case than in
the low-frequency approximation. The amplitude of $K_6$ is smaller
from the very beginning because the radiative metric variable $V_4$
absorbs some of the energy of the initial data and radiates it away in
a sharp pulse. This is not possible in the low-frequency
approximation, since there the radiation effects are neglected.

Finally, in Fig.~6, we show the dependence of the complex mode
frequency $\sigma$ on the polytropic index $n$, for a sequence of
polytropic models with fixed compactness $M/R = 0.2$ and a rotation
period of $1.0\,$ms.  We plot both the real and the imaginary part of
$\sigma$ in the same graph, using the left scale for the real part and
the right scale for the imaginary part. In addition we include the
region of the continuous spectrum, which is bounded by
\begin{equation}\label{range}
   m\Omega\(1 - \frac{2\varpi(0)}{\Omega l(l+1)}\) < \sigma_{cont}
  < m\Omega\(1 - \frac{2\varpi(R)}{\Omega l(l+1)}\)\;.
\end{equation}
For small $n$, the real part Re$(\sigma)$ lies outside the range of
the continuous spectrum, but for increasing $n$ it approaches the
lower boundary. The imaginary part Im$(\sigma)$ first grows with
increasing $n$, but eventually turns around and drops sharply down to
zero. In this case, Re$(\sigma)$ reaches the lower boundary of the
continuous spectrum and Im$(\sigma)$ goes to zero for $n = 0.843$. No
mode can be found for larger $n$.

\subsection{Realistic models}

As with the polytropic models, we can more or less take over the
results of paper I for models which are based on a realistic EOS.  It
is for the stiffest EOS that $r$-modes can exist for the whole
physically acceptable mass range, whereas for intermediate EOS, the
$r$-mode vanishes if the compactness exceeds a certain critical value.
For the softest EOS, none of the models can support $r$-modes.

\section{Summary}

We have performed both mode calculations and time evolutions of purely
axial perturbations of rotating neutron stars using the slow-rotation
approximation, i.e.~including only rotational correction terms linear
in the angular velocity $\Omega$. This work is thus an extension of
our previous results of Paper I. We have shown that on the one hand,
we are in full agreement with Newtonian-limit predictions for uniform
density models, but on the other hand, we obtain completely different
results for relativistic stars obeying polytropic equations of state
with the polytropic index $n$ being larger than about 0.8. In
agreement with our previous analysis of the $r$-modes in the
low-frequency approximation (Paper I), we showed that in these cases,
no discrete $r$-modes can be found.

For the uniform density models, we find that for less relativistic
models, the growth time is well described by a power law with an
exponent of $2l+2$, whereas for more relativistic models the strength
of the instability is somewhat suppressed. We conjecture that this can
be traced back to the appearance of a potential well inside the star,
which on the one hand can trap the gravitational waves and thus give
rise to very long-lived spacetime modes (the trapped modes), but on
the other hand should reduce the growth of the unstable $r$-mode.

In Paper I, we have shown that due to the singular structure of the
eigenvalue equation in the low-frequency approximation, there exists a
class of physically reasonable stellar models based on polytropic or
certain soft realistic equations of state, which does not permit the
existence of physical $r$-modes. As a result of the present work, it
became clear that even with the inclusion of the gravitational
radiation reaction, the slow-rotation approximation can still fail to
yield the $r$-mode solutions expected from Newtonian theory. Although
we can find very good agreement for uniform density models, the
$r$-modes seem to disappear for the relativistic polytropic cases with
large polytropic index $n$.

It might well be that inclusion of higher order corrections in
$\Omega$ or the coupling to the polar equations will drastically
change the above picture, and relativistic $r$-modes can be found for
any stellar model. However, it also might be the case that the above
picture persists, and that the $r$-modes simply do not exist for
certain stellar models, which might be a consequence of the
relativistic frame dragging. A first hint in this direction might come
from recent results by Karino et al.~(2001). They computed the
$r$-modes of rapidly and differentially rotating Newtonian polytropic
models without resorting to the slow-rotation approximation. They
found that for certain parameters of the rotation, their numerical
scheme ceases to yield mode solutions, because of the appearance of
co-rotation points. Through this analogy, we may conjecture that the
same could hold in the rapidly rotating case, where one does not
expand the equations in terms of $\Omega$.  Of course, not finding the
modes numerically does not prove their non-existence. However, the
results of Karino et al.~(2000) seem at least to indicate that even if
there exist $r$-modes in the regime where their code failed to detect
any, they might at least be strongly affected by the differential
rotation and have quite different physical properties than the
$r$-modes in weakly differentially or uniformly rotating stars.

The same could be true for the relativistic $r$-modes of those stellar
models where, within the slow-rotation approximation, one cannot find
any, as the relativistic frame dragging has an effect similar to that
of differential rotation. If it is only by including higher order
terms that the true physical $r$-modes can be found, we might expect
them to have quite different physical properties, and probably much
larger growth times, as would be expected from Newtonian theory. Of
course, if the true EOS of a neutron star is such that it already
permits the existence of $r$-modes in the slow-rotation approximation,
i.e.~if the EOS is rather stiff, and unless other physical mechanism
do not prevent the $r$-mode from growing, then the results obtained
from Newtonian theory should be reliable. On the other hand if the EOS
is rather soft, then it might well turn out that the implications of
the $r$-mode instability, as based on Newtonian estimates, such as
spin-down scenarios and gravitational wave emission, might be less
significant. Either way, a full understanding of the relativistic
$r$-modes is still outstanding and has to take into account at least
the coupling to the polar equations, if not even the inclusion of all
$\Omega^2$ corrections.

\section*{Acknowledgements}
We thank Nils Andersson, Horst Beyer, John Friedman, Ian Jones,
Luciano Rezzolla, Bernard F.~Schutz, Adamantios Stavridis, Nikolaos
Stergioulas and Shin Yoshida for helpful discussions. J.R.~is
supported by the Marie Curie Fellowship No.~HPMF-CT-1999-00364.
This work has been supported by the EU Programme 'Improving the Human
Research Potential and the Socio-Economic Knowledge Base' (Research
Training Network Contract HPRN-CT-2000-00137).

\label{lastpage}

\begin{figure}
\begin{center}
\vspace*{1cm}
\leavevmode
\begin{minipage}{8cm}
\epsfxsize=\textwidth
\epsfbox{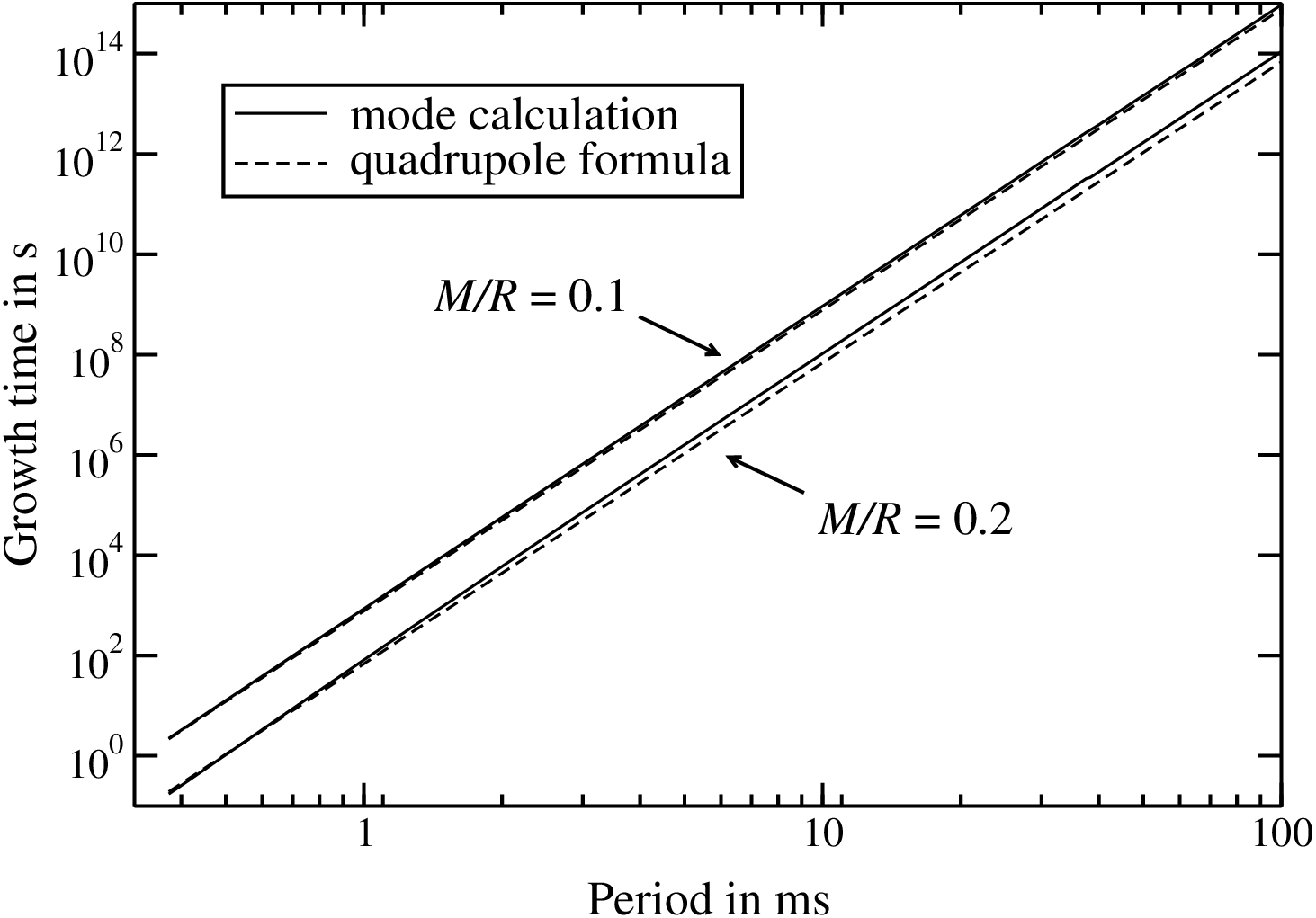}
\end{minipage}
\hspace*{5mm}
\begin{minipage}{8cm}
\epsfxsize=\textwidth
\epsfbox{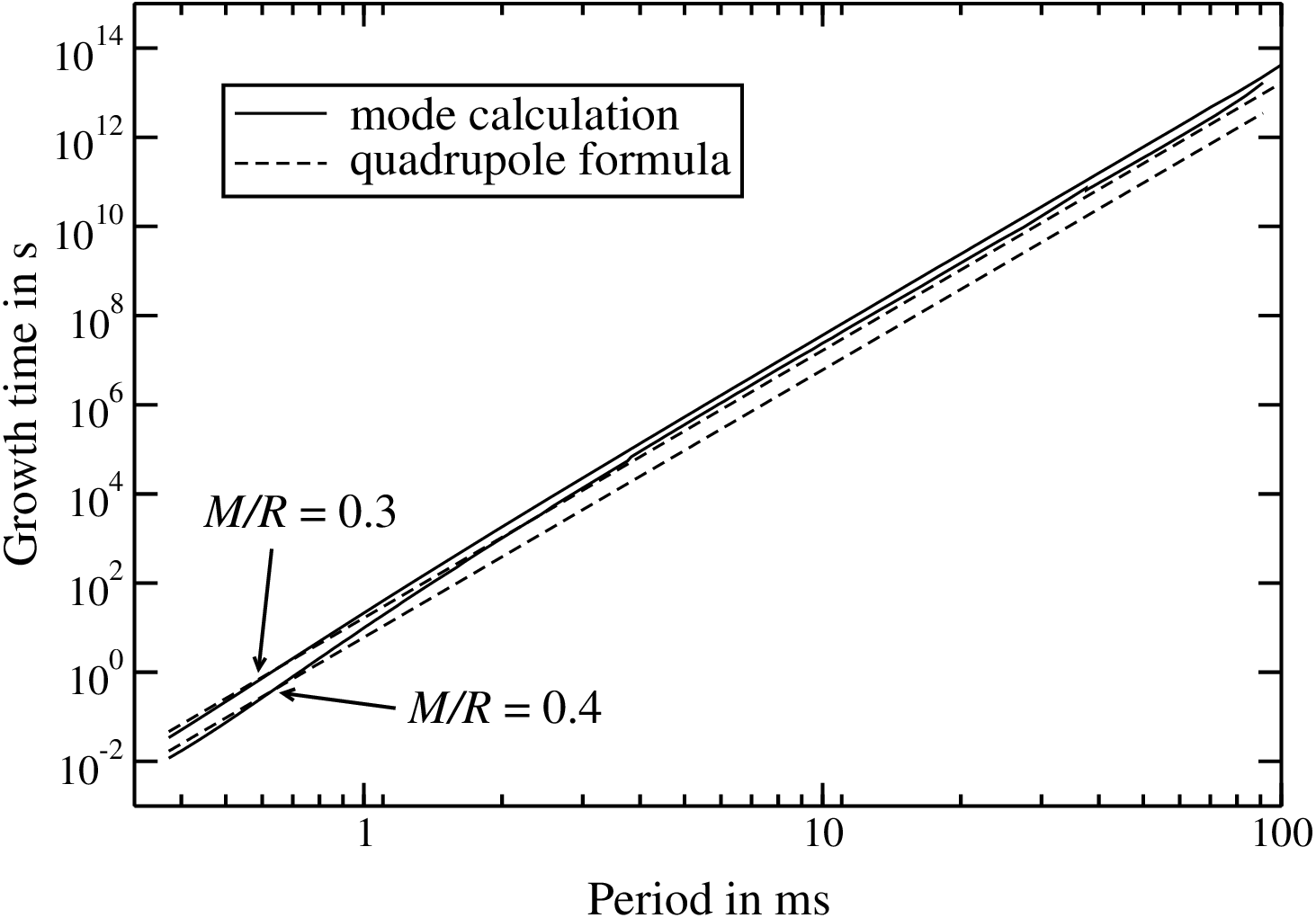}
\end{minipage}
\vspace*{5mm}
\caption{\label{fig1}Double logarithmic plot of the growth time for uniform
  density models with compactness $M/R$ = 0.1, 0.2 (left panel) and
  $M/R$ = 0.3, 0.4 (right panel). Also included is the growth time
  deduced from formula (\ref{g_rate}). For $M/R$ = 0.1 and 0.2, the
  agreement very good, whereas for $M/R$ = 0.3 and in particular for
  $M/R$ = 0.4, there is a systematic deviation, and formula
  (\ref{g_rate}) tends to overestimate the growth rate.}
\end{center}
\end{figure}

\begin{figure}
\begin{center}
\vspace*{1cm}
\leavevmode
\begin{minipage}{8cm}
\epsfxsize=\textwidth
\epsfbox{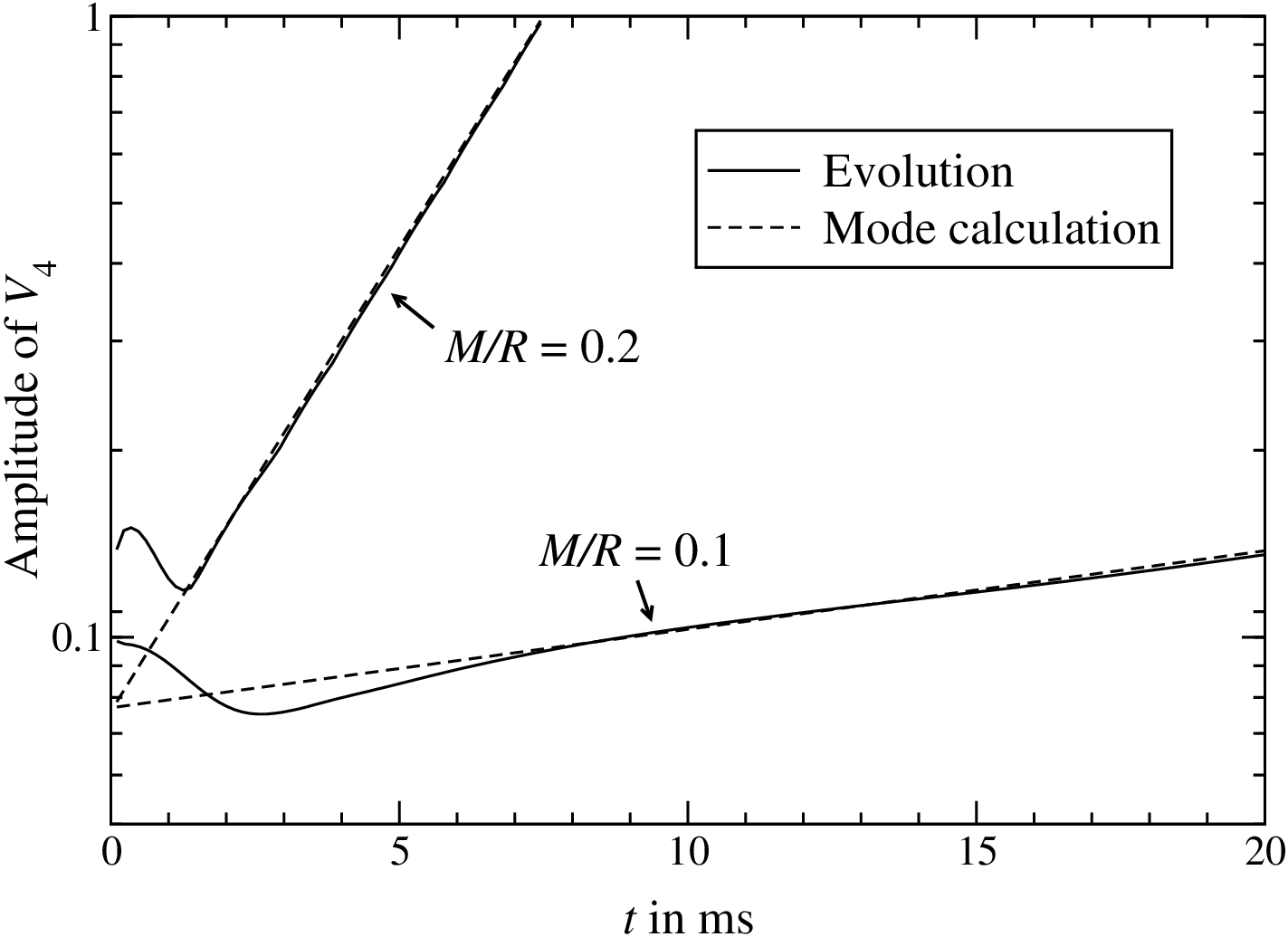}
\end{minipage}
\hspace*{5mm}
\begin{minipage}{8cm}
\epsfxsize=\textwidth
\epsfbox{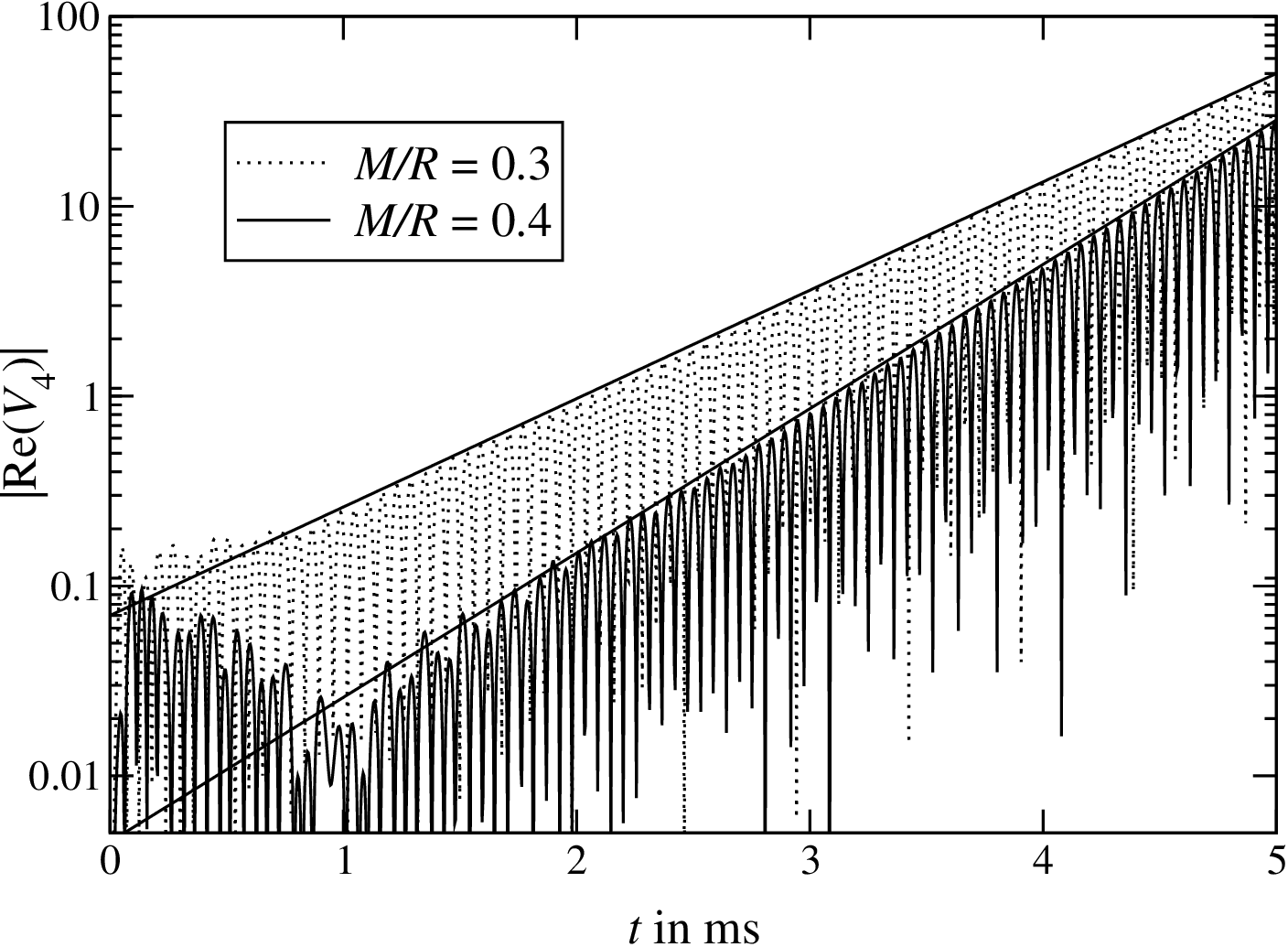}
\end{minipage}
\vspace*{5mm}
\caption{\label{fig2}Evolution of the 4 uniform density models of
  Table 1. Each model permits the existence of an $r$-mode, and the
  evolution shows the expected exponential growth in the amplitude.
  Also included are the slopes of the growth obtained from mode
  calculations. In the right panel, we have plotted only the amplitude
  of $V_4$ since the oscillations would be too dense.}
\end{center}
\end{figure}

\begin{figure}
\vspace*{1cm}
\begin{center}
\epsfxsize=8cm
\epsfbox{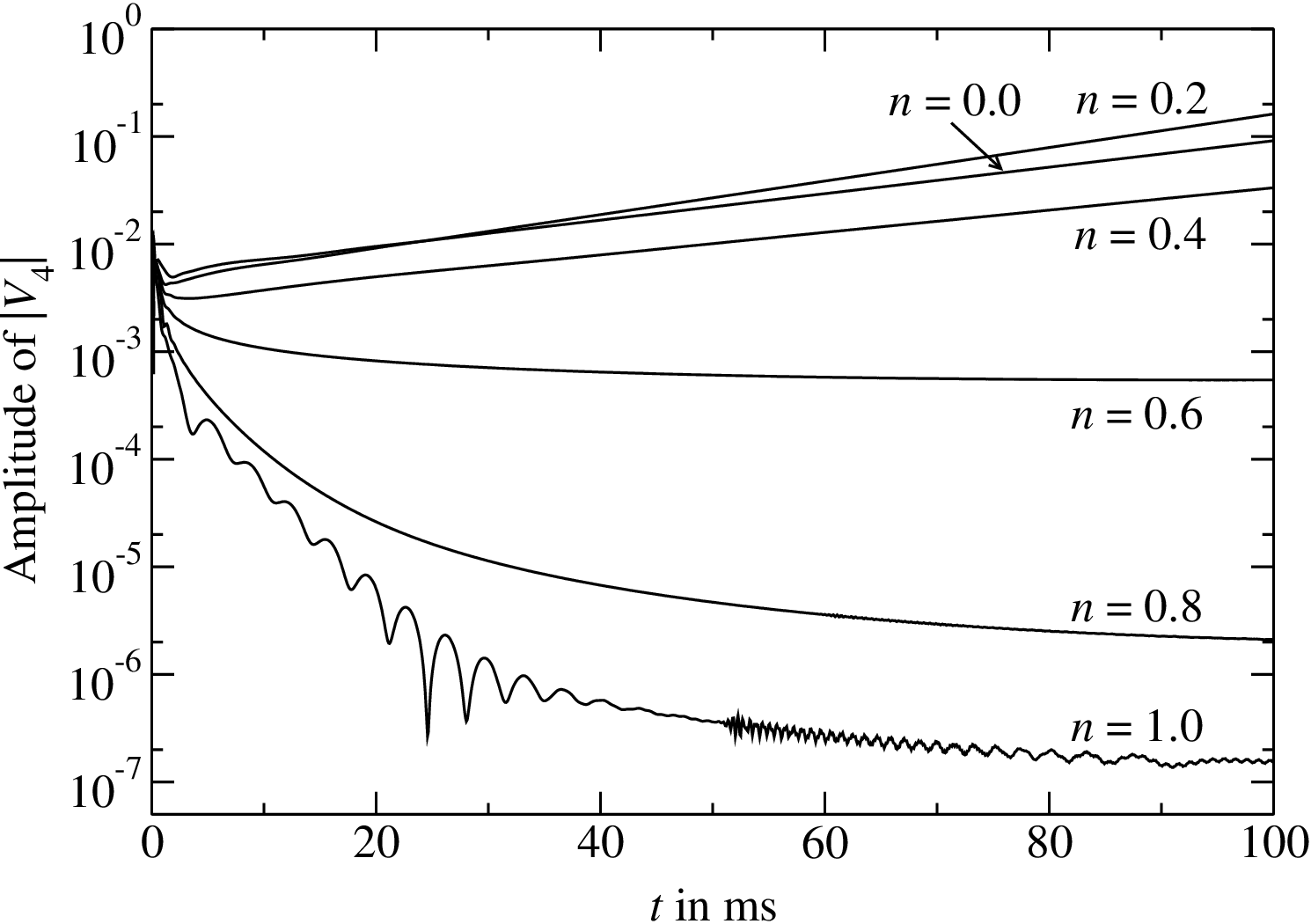}
\vspace*{5mm}
\caption{\label{fig3}Amplitude of $V_4$ for polytropic stellar models
  with the same compactness $M/R = 0.3$ and rotational parameter $\veps
  = 1.0$, but differing values of the polytropic index $n$, ranging
  from $n=0$ to $n=1$. For $n = 0$, 0.2 and 0.4, the evolutions show
  exponential growth, whereas for values of $n \ge 0.6$, the
  amplitudes decay.}
\end{center}
\end{figure}

\begin{figure}
\begin{center}
\vspace*{1cm}
\leavevmode
\begin{minipage}{8cm}
\epsfxsize=\textwidth
\epsfbox{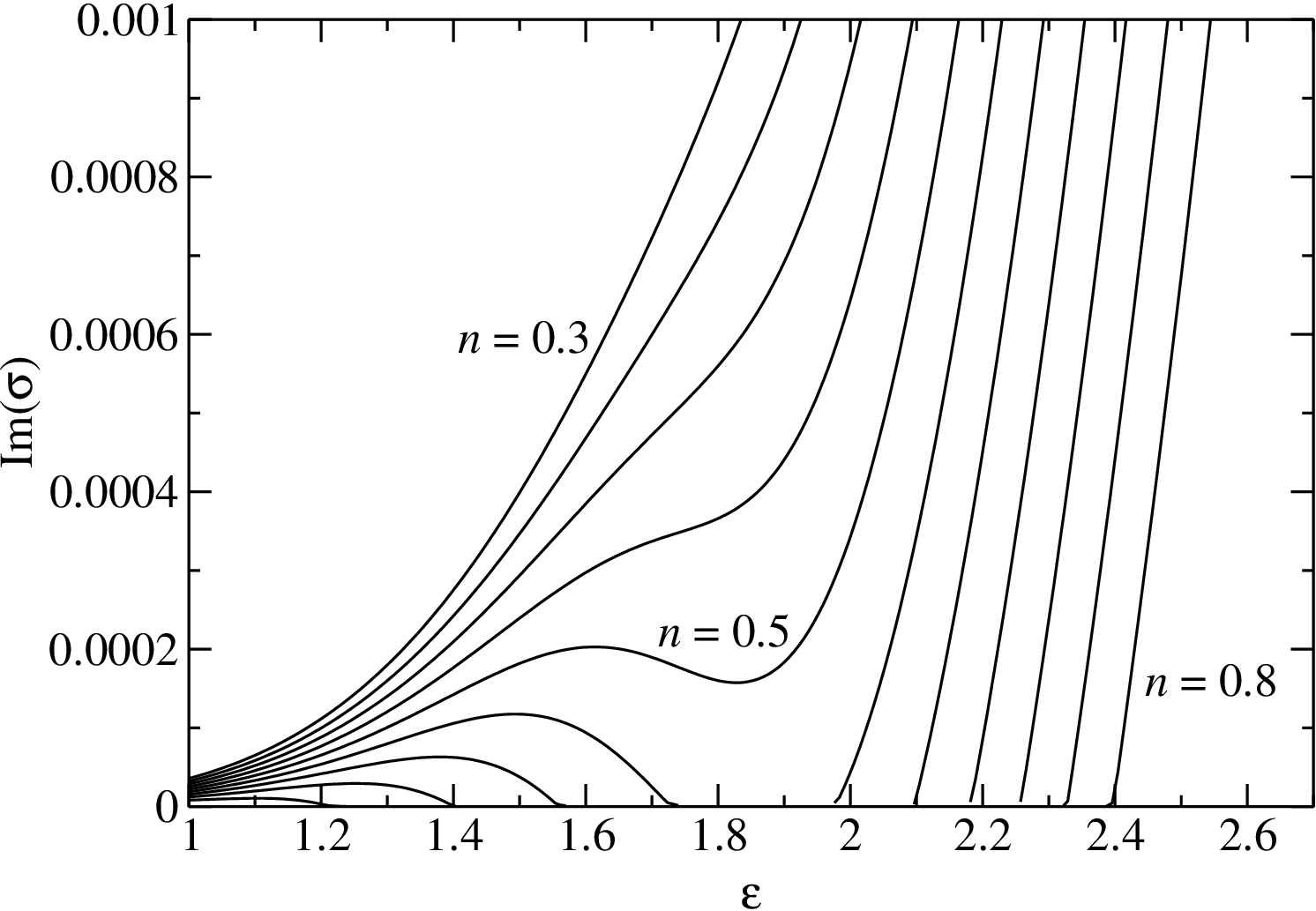}
\end{minipage}
\hspace*{5mm}
\begin{minipage}{8cm}
\epsfxsize=\textwidth
\epsfbox{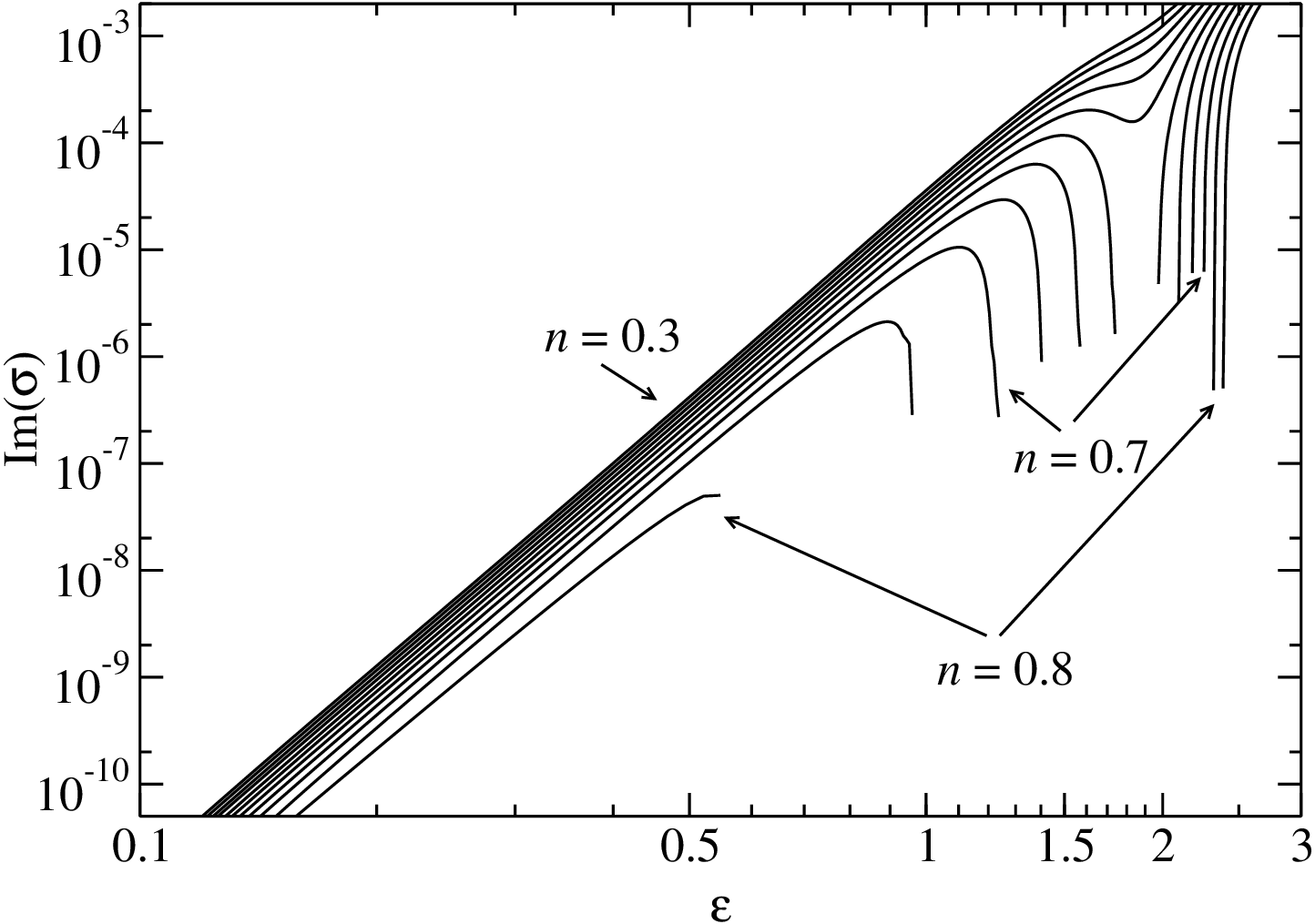}
\end{minipage}
\vspace*{5mm}
\caption{Growth times of the $r$-mode as a function of the rotational
  parameter $\veps$ for a sequence of polytropic stellar models with
  the polytropic index $n$ ranging from $n = 0.3$ to $n = 0.8$ in
  steps of 0.05. As $n$ is increased a deviation from the power law
  starts to emerge. Note that the values $\veps > 0.7$ are unphysical and
  are included only to gain a better understanding of the behaviour of
  the growth times for $\veps < 0.7$.}
\end{center}
\end{figure}

\begin{figure}
\vspace*{1cm}
\begin{center}
\epsfxsize=8cm \epsfbox{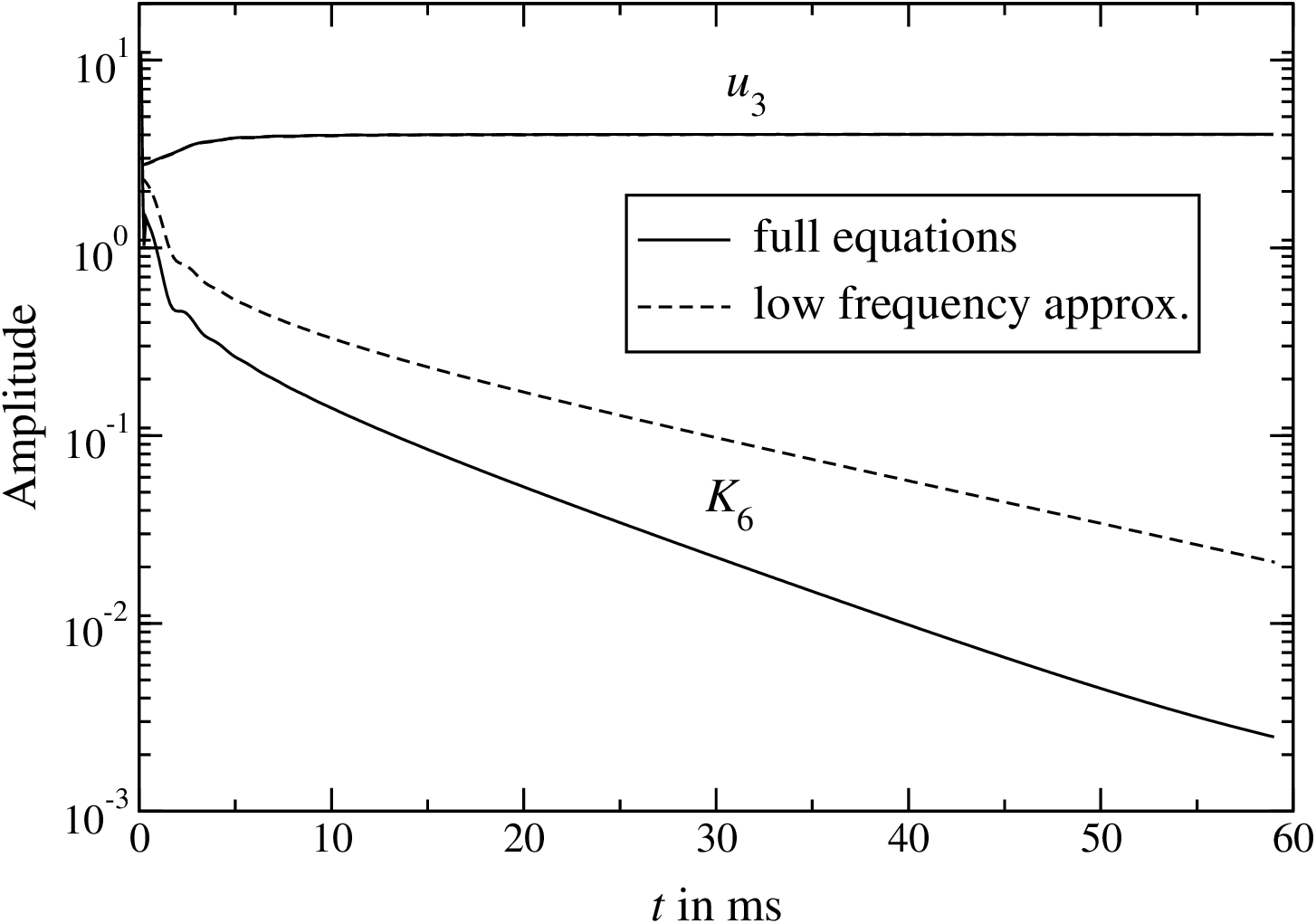} \vspace*{5mm}
\caption{\label{fig5}Comparison of the evolution of the same
  initial data in the full case and in the low-frequency approximation
  for a stellar model which does not possess a physical $r$-mode. It
  is shown the amplitude of $K_6$ outside the star together with the
  velocity perturbations $u_3$ extracted in the middle of the star. In
  both cases the fluid perturbation are almost identical and cannot be
  distinguished in the graph.  Furthermore, for the full equations,
  the decay of $K_6$ is even stronger than in the low-frequency
  approximation.}
\end{center}
\end{figure}

\begin{figure}
\begin{center}
\epsfxsize=8cm
\epsfbox{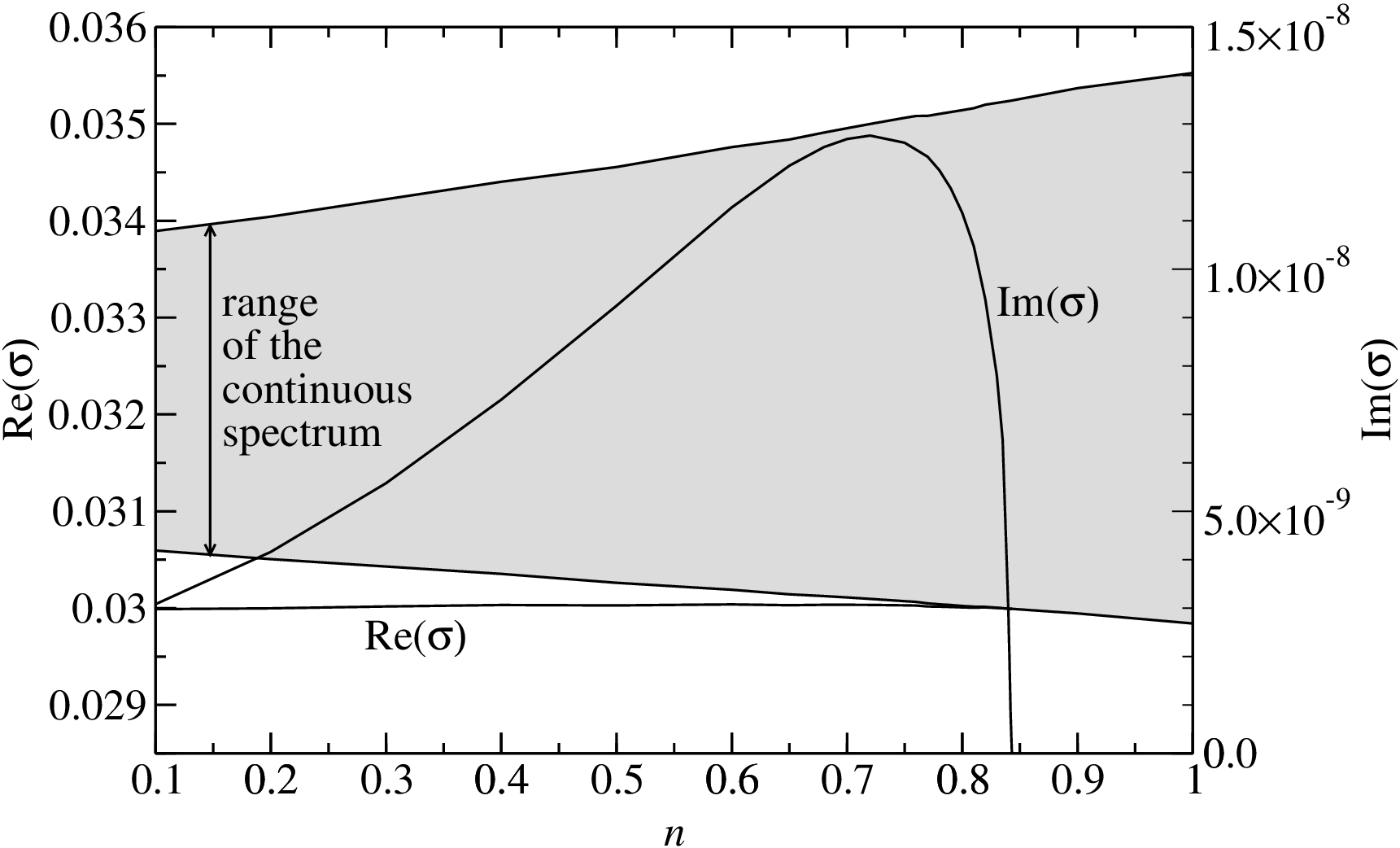}
\vspace*{5mm}
\caption{\label{fig6}Behaviour of the real and imaginary part of the
  $r$-mode frequency $\sigma$ as a function of the polytropic index
  $n$ for models with $M/R = 0.2$ and rotation period of $1.0\,$ms.
  The scale on the left is for the real part, the scale on the right
  for the imaginary part of $\sigma$. For small $n$, the real part
  Re$(\sigma)$ lies outside the range of the continuous spectrum.  For
  increasing $n$, Re$(\sigma)$ approaches the lower boundary and the
  imaginary part Im$(\sigma)$ eventually starts to decrease.  For $n =
  0.843$, $\Re(\sigma)$ reaches the lower boundary of the continuous
  spectrum and Im$(\sigma)$ goes to zero. For larger $n$, no mode can
  be found any more.}
\end{center}
\end{figure}

\begin{thebibliography}{99}

\bibitem{And98} Andersson N., 1998, ApJ, 502, 708

\bibitem{AK00} Andersson N., Kokkotas K.D., 2001, Int. J. Mod. Phys. D,
in press; gr-qc/0010102

\bibitem{ADM} Arnowitt R., Deser S., Misner C.W., 1962, in Witten L., ed.,
  Gravitation: An Introduction to Current Research.
  Wiley, New York, p.227

\bibitem{BK99} Beyer H.R., Kokkotas K.D., 1999, MNRAS, 308, 745

\bibitem{FL01} Friedman J.L., Lockitch K.H., 2001, gr-qc/0102114
  
\bibitem{KYE01} Karino S., Yoshida S., Eriguchi Y., 2001, Phys. Rev. D, 64, 024003 

\bibitem{Koj92} Kojima Y., 1992, Phys. Rev. D, 46, 4289

\bibitem{Koj97} Kojima Y., 1997, Prog. Theor. Phys. Suppl., 128, 251

\bibitem{Koj98} Kojima Y., 1998, MNRAS, 293, 49

\bibitem{KDK94} Kokkotas K.D., 1994, MNRAS, 268, 1015; Erratum:
1995, 277, 1599

\bibitem{KS99} Kokkotas K.D., Stergioulas N., 1999, A\&A , 341, 110

\bibitem{LAF01} Lockitch K.H., Andersson N., Friedman J.L., 2001,
Phys. Rev. D, 63, 024019

\bibitem{Ruoff01} Ruoff J., 2001, Phys. Rev. D, 63, 064018

\bibitem{RK01} Ruoff J., Kokkotas K.D., 2001, gr-qc/0101105


\end{thebibliography}
\end{document}